\documentclass[a4paper]{jpconf}
\usepackage[justification=justified,
            singlelinecheck=false,
            position=top]{caption}

\usepackage{hyperref}
\usepackage{graphicx}  
\usepackage{amssymb}   
\usepackage{bbm}
\usepackage{dsfont}
\usepackage{amsfonts}
\usepackage{epsfig}
\usepackage{amsmath}
\usepackage{color}
\usepackage{xcolor}

\def\bEq#1\eEq{\begin{align}#1\end{align}} 

\newcommand{\Fig}[1]{Fig.~\ref{#1}}
\newcommand{\Eq}[1]{Eq.~(\ref{#1})}
\newcommand{\Ref}[1]{Ref.~\cite{#1}}

\newcommand{\dx}[1]{ \mathrm{d}{#1} \,}

\newcommand{\kk}[1]{ {\mathbf{k}_{#1}} }

\newcommand{\rtw}{\rightarrow}

\newcommand{\nt}{\notag}
\newcommand{\ve}{{\varepsilon}}
\newcommand{\vrh}{{\varrho}}

\newcommand{\EF}{{\varepsilon_F}}
\newcommand{\KF}{{k_F}}
\newcommand{\I}{{\mathrm{i}}}
\newcommand{\NF}{{n_F}}

\renewcommand{\Im}{{\mathrm{Im}\,}}

\usepackage{letltxmacro}
\makeatletter
\let\oldr@@t\r@@t
\def\r@@t#1#2{%
\setbox0=\hbox{$\oldr@@t#1{#2\,}$}\dimen0=\ht0
\advance\dimen0-0.2\ht0
\setbox2=\hbox{\vrule height\ht0 depth -\dimen0}%
{\box0\lower0.4pt\box2}}
\LetLtxMacro{\oldsqrt}{\sqrt}
\renewcommand*{\sqrt}[2][\ ]{\oldsqrt[#1]{#2}}

\makeatletter
\newsavebox{\@brx}
\newcommand{\llangle}[1][]{\savebox{\@brx}{\(\m@th{#1\langle}\)}%
  \mathopen{\copy\@brx\mkern2mu\kern-0.9\wd\@brx\usebox{\@brx}}}
\newcommand{\rrangle}[1][]{\savebox{\@brx}{\(\m@th{#1\rangle}\)}%
  \mathclose{\copy\@brx\mkern2mu\kern-0.9\wd\@brx\usebox{\@brx}}}
\makeatother

\newif\ifContLineOne
\newif\ifContLineTwo
\newif\ifContLineThree

\def\conC#1{\vbox{\ialign{##\crcr
  \ifContLineThree\hrulefill\else\vphantom{\hrulefill}\fi\crcr
  \noalign{\kern3.2pt\nointerlineskip}
  \ifContLineTwo\hrulefill\else\vphantom{\hrulefill}\fi\crcr
  \noalign{\kern3.2pt\nointerlineskip}
  \ifContLineOne\hrulefill\else\vphantom{\hrulefill}\fi\crcr
  \noalign{\nointerlineskip}
  $\hfil\textstyle{\vbox to 14pt{}#1}\hfil$\crcr}}}

\def\DrawLeg#1#2{
  \kern-.2pt              
  \dimen2 =#1             
  \advance\dimen2 by 2pt  
  \dimen3 = 10.6pt        
  \dimen4 =3.6pt          
  \advance\dimen3 by -\dimen2 
  \multiply\dimen4 by #2
  \advance\dimen3 by \dimen4
  \raise\dimen2 \hbox{\vrule height\dimen3 width .4pt} 
  \kern-.2pt}             

\def\begC#1#2{\setbox0 =\hbox{$\textstyle{#2}$}
  \dimen0=.5\wd0 \dimen1=\ht0
  \conC{\hskip\dimen0}
  \count255=#1
  \ifnum\count255 =1 \ContLineOnetrue\else
  \ifnum\count255 =2 \ContLineTwotrue\else
  \ifnum\count255 =3 \ContLineThreetrue\fi\fi\fi
  \DrawLeg{\dimen1}{\count255}
  \conC{\hskip\dimen0}
  \kern-\dimen0\kern-\dimen0 \box0}

\def\endC#1#2{\setbox0 =\hbox{$\textstyle{#2}$}
  \dimen0=.5\wd0 \dimen1=\ht0
  \conC{\hskip\dimen0}
  \count255=#1
  \ifnum\count255 =1 \ContLineOnefalse\else
  \ifnum\count255 =2 \ContLineTwofalse\else
  \ifnum\count255 =3 \ContLineThreefalse\fi\fi\fi
  \DrawLeg{\dimen1}{\count255}
  \conC{\hskip\dimen0}
  \kern-\dimen0\kern-\dimen0 \box0}

\graphicspath{%
  {./figures/}%
}

\begin{document}
\title{Oscillation and suppression of Kondo temperature by RKKY 
coupling in two-site Kondo systems
}

\author{Ammar Nejati$^1$ and Johann Kroha$^{1,2}$}

\address{
$^1$ Physikalisches Institut and Bethe Center for Theoretical Physics,
Universit{\"a}t Bonn, Nussallee~12, D-53115 Bonn, Germany}
\address{
$^2$ Center for Correlated Matter, Zhejiang University, 
Hangzhou, Zhejiang 310058, China
}

\ead{kroha@th.physik.uni-bonn.de}

\begin{abstract}
We apply our recently developed, selfconsistent renormalization group (RG)
method~\cite{Nejati16} to STM spectra of a two-impurity Kondo system 
consisting of two cobalt atoms connected by a one-dimensional Cu 
chain on a Cu surface~\cite{Neel11}. This RG method was developed to describe 
local spin screening in multi-impurity Kondo systems in presence of the 
Ruderman-Kittel-Kasuya-Yosida
(RKKY) interaction. Using the RKKY interaction of a one-dimensional chain, 
we explain the experimentally observed suppression and oscillation of the 
Kondo temperature, $T_K(y)$, as a function of the length of the chain 
and the corresponding RKKY interaction parameter $y$, 
regardless of the RKKY coupling being ferromagnetic or 
antiferromagnetic.
\end{abstract}

\section{Introduction}
\label{sec:introduction}

The Kondo effect is a genuine many-body phenomenon which, in its usual
magnetic manifestation, appears when a magnetic ion is immersed in a
non-magnetic host metal. At high temperatures, the magnetic moment of the ion
exhibits a Curie magnetic susceptibility corresponding to uncorrelated moments
in a paramagnetic phase. When the temperature is decreased below a 
characteristic scale, the Kondo temperature $T_K$, the local moment is 
collectively screened by the conduction electron spins, leading to a 
narrow resonance in the electronic spectrum and a temperature-independent 
Pauli susceptibility~\cite{Hewson97,Coleman07}.
The Kondo effect is observed in a variety of systems, ranging from metallic
alloys~\cite{deHaas34,Ehm07} to artifical semiconductor nanostructures
like quantum dots~\cite{GoldhaberGordon98,Weis98,Kouwenhoven00,Grobis07} 
and molecular devices~\cite{Nygaard00,Park02,Scott10}.

In a spin-1/2 two-impurity Kondo (2IK) system, 
two different ground states are possible, the Kondo state where each of the
two local spins forms a Kondo singlet with the conduction electrons, and a 
dimer state where the two local spins are mutually bound into a singlet 
or triplet. The two states are distinguishable by different behavior of
the spin correlation functions between a local spin and the conduction 
electrons on one hand and between the local spins among each other on the
other hand. It has been recognized \cite{Nejati16} that it makes 
a crucial difference whether there is a direct exchange coupling between 
the impurities or whether the interimpurity coupling is mediated by the 
conduction electrons, the Ruderman-Kittel-Kasuya-Yosida (RKKY) interaction 
\cite{Ruderman54,Kasuya56,Yosida57,vVleck62}. 
With a direct exchange coupling, a quantum phase transition 
between the Kondo and the dimer phase \cite{Jones88,Jones89}
occurs only for a particular particle-hole symmetry \cite{Affleck95}, 
associated with a coupling-decoupling transition of the conduction band. 
In the case of RKKY-mediated interimpurity coupling, a complete decoupling of 
the conduction electrons at low energy is not possible. However, it 
was recently predicted that nevertheless a breakdown of the complete 
Kondo screening occurs at a critical strength of a dimensionless 
RKKY coupling parameter $y$ \cite{Nejati16}, associated with a transition 
to a phase of partially screened local moments and a universal reduction 
of the Kondo scale $T_K(y)$. Numerical renormalization group (NRG) studies
have recently been performed also for indirect interimpurity coupling
\cite{Mitchell15}.

2IK systems have been realized experimentally in semiconductor 
quantum dot devices~\cite{Craig04,Sasaki06,Jeong01,Chang09} and in systems 
of magnetic adatoms on metal surfaces, studied by 
scanning tunneling spectroscopy~\cite{Neel11,Chen99,Bork11,Prueser14}. 
In the present work we focus on understanding the results of \Ref{Neel11}.
In this experiment, N\'eel {\it et al.} have assembled atomic chains of 
up to $n=6$ Cu atoms on a Cu surface, flanked by a magnetic Co atom
on each end. This forms a 2IK system where the indirect RKKY interaction is 
enhanced by the quasi-one-dimensional (1D) Cu chain. 
The Kondo temperature was
deduced from the width of the Kondo-Fano resonance~\cite{Uhsaghy00} 
in the $dI/dV$ signal taken above one of the two Co atoms. It shows 
an oscillatory suppression as function of the separation of the Co atoms.

This paper is organized as follows: In section~\ref{sec:RKKY}, we
derive the form of the RKKY interaction for the 1D chain and use the
result in section~\ref{sec:RKKYRG} to obtain the Kondo temperature via the 
RKKY-modified RG. In the final section~\ref{sec:results}, we provide a detailed
comparison of the results with the experimental observations.

\section{RKKY interaction mediated by the 1D chain}
\label{sec:RKKY}

The two-impurity Kondo system is described by the Hamiltonian,
\bEq
\label{eq:2impkondoH}
H_{\text{2IK}} = \sum_{\kk{} \sigma} \ve_{\kk{} \sigma} \, c_{\kk{}
  \sigma}^\dagger c_{\kk{} \sigma} 
+ J_0 \sum_{i = 1,2} \mathbf{S}(x_i) \cdot \mathbf{s}(x_i) ~,
\eEq
where $\mathbf{S}(x_i)$ and $\frac{1}{2} \mathbf{s}(x_i)$ 
represent the spin operators of the impurities and conduction electrons 
at the positions of the impurities, $x_{1,2}$, respectively.
($\kk,\sigma$) represents a complete basis of the conduction electron system,
composed of the substrate and the 1D Cu chain. Density functional theory
(DFT) calculations indicate \cite{Neel11} that the RKKY coupling is 
dominated by the 1D chain (because in the 3D substrate the RKKY interaction 
decays faster than in 1D) and that the local DOS in the Cu chain is well 
approximated by an infinitely long chain. Deviations from this behavior 
are found only for the shortest chains of length $n=1$, 
see section \ref{sec:results}. We, therefore, approximate the 
conduction-mediated coupling between the Co spins at positions $x_1$ 
and $x_2$ by the RKKY coupling of two spins separated by $x=x_2-x_1$
on an infinite chain (infinite chain approximation). 
In a simple model for the Cu chain we assume a Fermi momentum of
$k_F=\pi/(2a)$ corresponding to half filling of the Cu 4s band,
where $a$ is the Cu lattice spacing.

\begin{figure}
    \centering
    \includegraphics[width=0.95\linewidth]{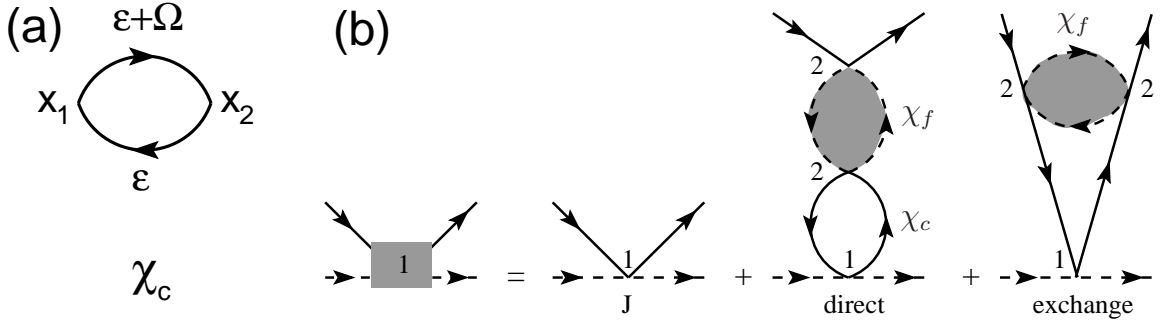}
\vspace*{0.4cm}
  \caption{(a) Conduction electron susceptibility or ``polarization bubble''. 
    (b) Total conduction electron-impurity spin vertex, including
    non-local RKKY vertex corrections (`direct' and `exchange' diagrams). 
    The solid (dashed) lines represent conduction electron (impurity spin)
    propagators, respectively. 
    $ \chi_c $ ($ \chi_f $) is the conduction electron (impurity) 
    susceptibility. 
    \label{fig:diagrams}
  }
\end{figure}

In leading order perturbation theory in the local spin exchange coupling
$J_0$, the RKKY coupling reads, 
$I_{\text{RKKY}} (x) = - \chi_c (x)$, where $\chi_c (x)$ is the free
conduction electron susceptibility (or polarization) 
of the chain~\cite{Aristov97}, as shown in \Fig{fig:diagrams} (a). 
We calculate $\chi_c (x)$ here directly in position space, 
as this will be needed for the RG vertex in section \ref{sec:RKKYRG} 
and is more concise than the momentum space calculation \cite{Aristov97}.    
\begin{eqnarray}
\label{eq:csusc}
\chi_c^R (x_2, x_1 ; \Omega) =  \int_{-\infty}^{+\infty} \dx{\varepsilon} \, 
\NF(-\varepsilon) \,
\Big( G_c^R(x ; \varepsilon + \Omega) \, A_c(- x; \varepsilon)  
+ A_c(x; \varepsilon) \, G_c^A(-x; \varepsilon - \Omega) \Big) ~,
\end{eqnarray}
where $ G_c^{R/A} $ is the retarded/advanced Green's function for the free
electrons and $A_c$ represents their spectral function. $\NF(\varepsilon)$ 
denotes the Fermi-Dirac distribution, and $\varepsilon =  E - \EF $ 
is the energy with respect to the Fermi level, $\EF$.
The static susceptibility entering the RG vertex \cite{Nejati16} 
is obtained from the zero-frequency limit 
$\Omega \rtw 0 $, and $\NF(\varepsilon)=\Theta(\varepsilon)$ for 
temperature $T=0$. Assuming, for simplicity, a quadratic dispersion,
$\ve_k = k^2/2m$, the 1D free Green's function (per spin orientation) 
reads \cite{Economou06},
\bEq
\label{eq:cgreenf}
G_c^{0, R/A} (x, \omega) &= \mp\I \pi {\cal N}(\omega) \exp[ \pm\I k(\omega) | x | ] ~, \nt \\
A_c(x, \omega) &= -\frac{1}{\pi} \Im G_c^{0, R} (x, \omega) =  
{\cal N}(\omega) \cos[ k(\omega) |x| ] ~,
\eEq
where ${\cal N}(\omega)=m/(2\pi k(\omega))$ is the continuum conduction 
electron DOS and 
$ k(\omega) = \sqrt{ 2 m (\EF +\omega) }$ the momentum corresponding to the 
energy $\omega$.  
Hence, the static conduction electron susceptibility is real and reads, 
\bEq
\chi_c^R (x_2, x_1 ; \Omega = 0) &\equiv  \chi_c^R (\vrh ; \Omega = 0) =
-\frac{1}{2} {\cal N}(0)  \, \left( \mathrm{Si}(\vrh) - \frac{\pi}{2} \right) ~,
\eEq
with $ \vrh = 2\KF |x| $, and  
$ \mathrm{Si}(z) = \int_{0}^{z} \dx{x} \frac{\sin(x)}{x}$ the integral sine.
Note that the long-distance limit of $\chi_c^R (x ; \Omega = 0)$ is
\bEq
\label{eq:csusc_asymp}
\chi_c^R (x ; \Omega = 0)\,\, \stackrel{|x|\to\infty}{\longrightarrow}\,\, \sim
\frac{\cos(\vrh)}{\vrh} ~,
\eEq
which manifests its slow decay and oscillatory behavior as a function of
interimpurity spacing.


\section{Renormalization group analysis for symmetrically coupled 
impurities}
\label{sec:RKKYRG}

In the previous section, we calculated the RKKY interaction
between the two Co impurities mediated by the 1D Cu chain connecting the 
magnetic ions. The RKKY interaction implies that the total spin exchange
vertex between conduction electrons and a local impurity spin acquires 
nonlocal contributions: A conduction electron can scatter from impurity 
2, and that spin flip can be transfered to impurity 1 by the itinerant
conduction electrons travelling along the Cu chain from the 2nd to the 1st 
Co impurity.  This process is depicted diagrammatically 
in \Fig{fig:diagrams} (b). As seen from the figure, this transfer process
necessarily involves the total, local spin susceptibility 
$\chi_f$ of the 2nd impurity. Its exact behavior is known from the 
Bethe ansatz solution \cite{Andrei83}. At low temperatures, $T < T_K$,
it is proportional to $1/T_K$ and crosses over to Curie behavior
$\chi_f\sim 1/T$ for $T\gg T_K$.

\begin{figure}
  \begin{minipage}{.42\textwidth}
    \includegraphics[width=1.0\linewidth]{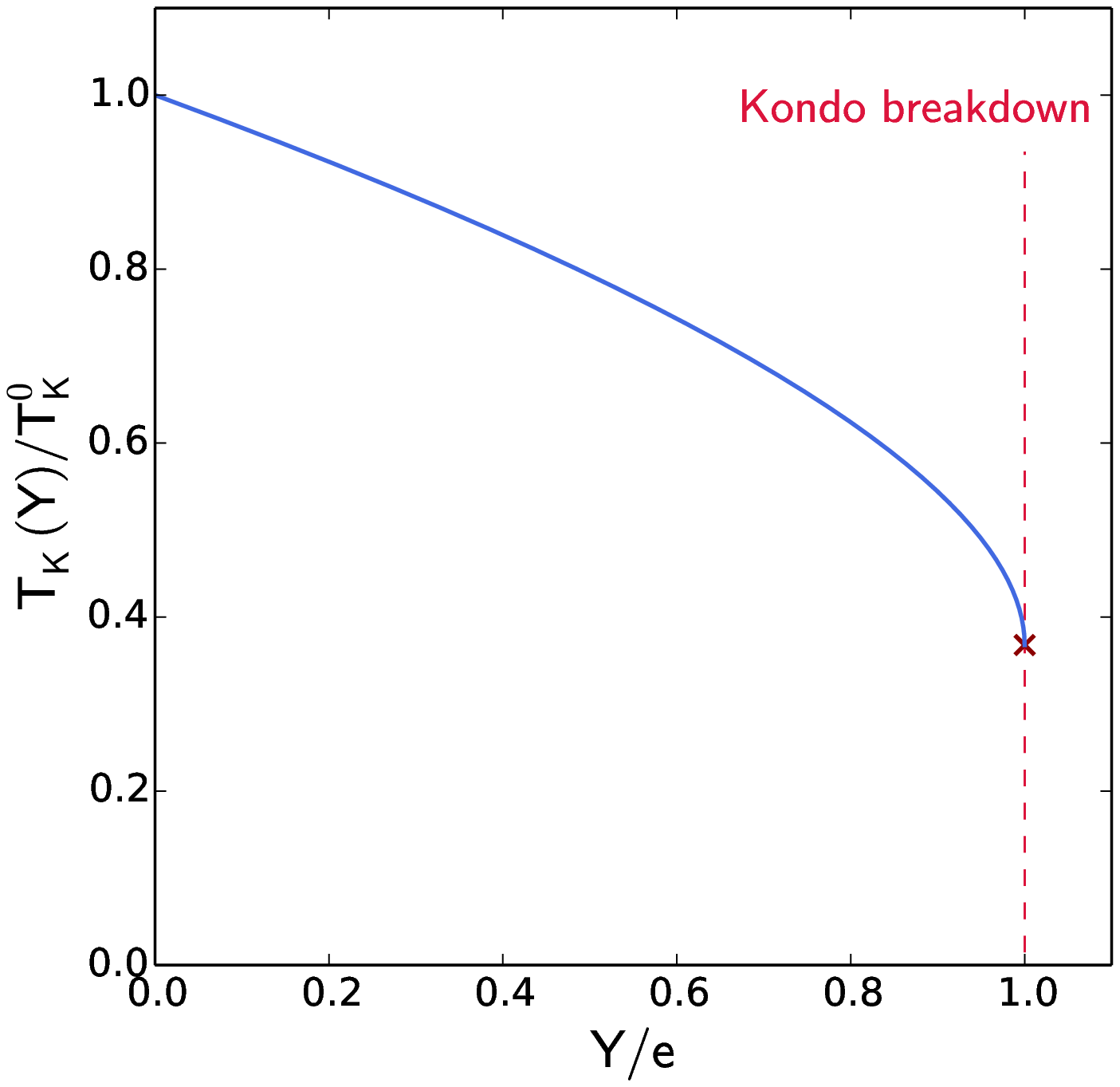}
  \end{minipage}
  \hspace*{0.000\textwidth}
  \begin{minipage}{.58\textwidth}
    \includegraphics[width=1.0\linewidth]{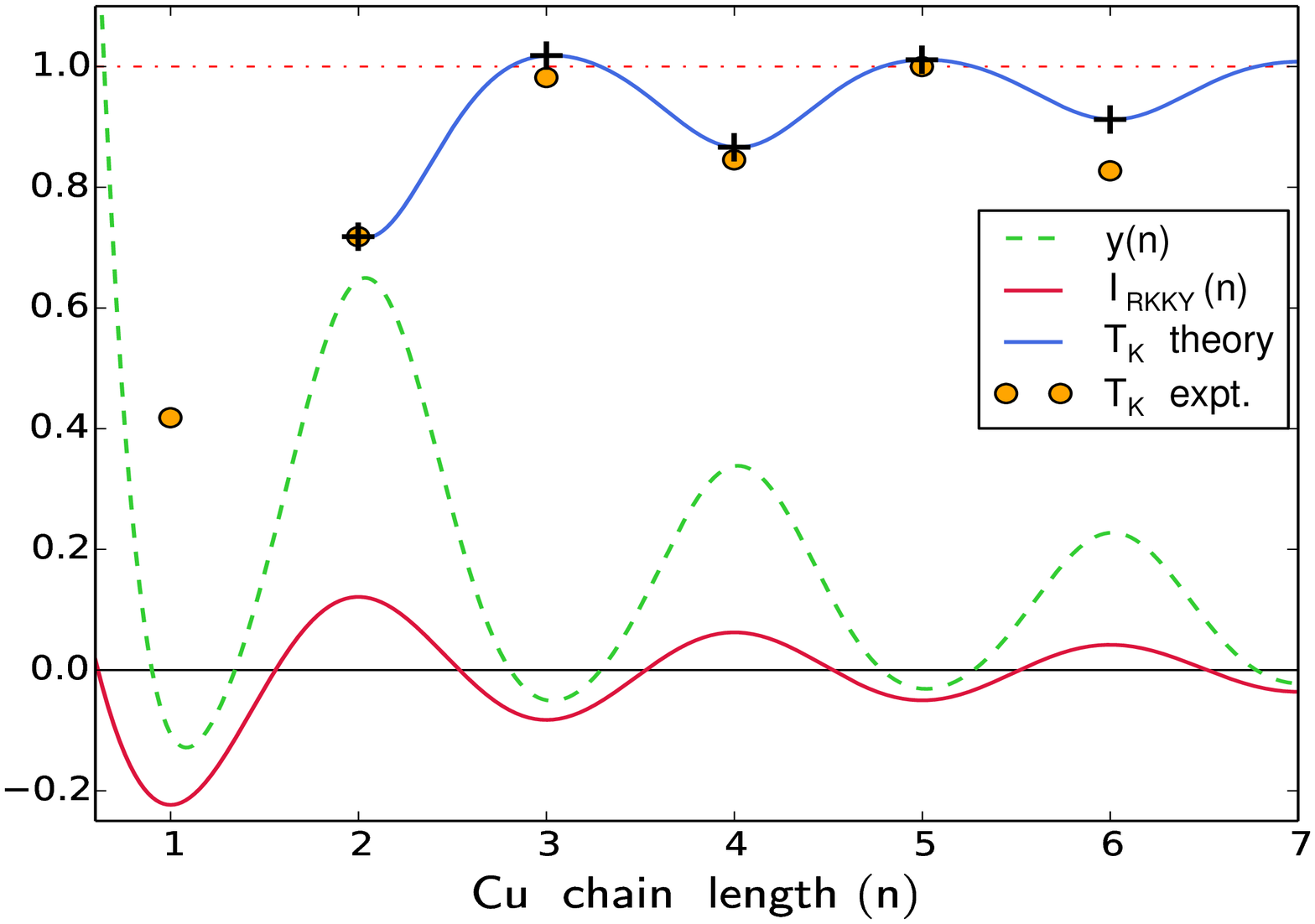}
  \end{minipage}
  \vspace*{0.4cm}
  \caption{{\it Left panel:} 
    The universal curve for the RKKY-modified Kondo temperature, 
    $T_K(y)/T_K(0)~=~\frac{-y/({\rm e}y_c)}{W(-y/({\rm e}y_c))}$, 
    as a function of the normalized RKKY strength parameter, 
    $ y/y_c $, where the critical RKKY parameter for Kondo breakdown 
    is $y_c = 1/({\rm e}\gamma_0)$ and ${\rm e}\approx 2.718$ is Euler's 
    constant (see \Eq{eq:TKsol}). Complete Kondo singlet formation ceases 
    to exist for $y>y_c$. 
    {\it Right panel:} 
    Oscillation and suppression of the Kondo temperature, $T_K(y(n))/T_K(0)$, 
    for Cu chains of length $n$: Comparison of theory and experiment
    \Ref{Neel11}. In the experiment, the single-ion Kondo scale 
    is $T_K(0)=110$~K. The overall amplitude of $T_K(y(n))/T_K(0)$ 
    is obtained by scaling the theoretical value to the experimental 
    one for $n=2$. It is remarkable that the 1D infinite chain model 
    accurately reproduces the experimantal data even for short chains 
    of $n\geq 2$. 
    The RKKY coupling, $I_{\mathrm{RKKY}}(n)$ (red solid line) and the 
    RKKY strength parameter, $y(n)$, averaged over one lattice spacing $a$ 
    (dashed green line), are also shown (arbitrary units).
  }
  \label{fig:TKy}
\end{figure}

With the expression for the RKKY-modified Kondo vertex, 
the RG equations for the Kondo coupling can be obtained by the
same procedure as developed in~\Ref{Nejati16}.
The RG equation for the dimensionless Kondo coupling, $g=J{\cal N}(0)$, 
as a function of the running conduction bandwidth, $D$, is obtained 
as \cite{Nejati16}, 
\bEq
\label{eq:RG_eq}
\frac{d g}{d \ln D} &= -2 g^2 \left( 1 - y \, g_{0}^2 \, \frac{D_0}{T_K} \frac{1}{\sqrt{1 + (\frac{D}{T_K})^2}} \right) ~,
\eEq
where $D_0$ is the full bandwidth of the conduction electrons, 
$g_0= J_0 {\cal N}(0)$  the dimensionless, 
bare Kondo coupling, $T_K = T_K(y) $ is the
RKKY-modified Kondo scale (to be determined), 
and $y$ is a dimensionless RKKY strength parameter. It is obtained for a 
2IK system as (c.f. \Ref{Nejati16}),
\begin{eqnarray}
y=-\frac{8W}{\pi ^2} {\rm Im} 
\left[
\frac{{\rm e}^{-\I k_Fx}}{N(0)^2}
 G_c^R(x, \Omega=0 ) \,
\chi_c(\vrh, \Omega=0 )
\right] = \frac{8W}{\pi {\cal N}(0)} \,\cos(\vrh)\,\chi_c(\vrh, \Omega=0 ) ~,
\label{eq:RKKYparameter}
\end{eqnarray}
with $W=2$ the Wilson ratio of a single Kondo impurity.
Note that the term proportional to $ \frac{1}{T_K} $ on the right-hand side
of~\Eq{eq:RG_eq} originates from the behavior of the local impurity
susceptibility, $ \chi_f \propto \frac{1}{T_K} $~\cite{Andrei83}. 
Furthermore, note that the RKKY parameter $y$ of the 2IK system is 
generically positive, even though $\chi_c^R(x,0)$ has an 
alternating sign, as shown in \Fig{fig:TKy}, left panel. 
Technically, this is because 
in $y$ the oscillatory factors 
$\exp(-\I k_Fx)\,G_c^R(x, \Omega=0 )$ and 
$\chi_c(\vrh, \Omega=0 )$ combine to form an essentially positive term.  
It is physically expected because the interimpurity RKKY coupling should 
reduce the onsite Kondo spin fluctuations irrespective of the RKKY coupling
being ferro- or antiferromagnetic.

The RG equation (\ref{eq:RG_eq}) is straight-forwardly integrated by
separation of variables,
\bEq
\int_{g(D_0)}^{g (D)} \frac{\dx{g}}{-2 g^2} =
\int_{D_0}^{D} \frac{\dx{D}}{D} \left( 1 - y g_{0}^2 \frac{D_0}{T_K} \frac{1}{\sqrt{1 + (D/T_K)^2} } \right) ~.
\eEq
To deduce the Kondo temperature $T_K$, one should use the fact that, 
as the running cutoff reaches the Kondo scale $D\to T_K$, the effective 
exchange coupling diverges, $g\to \infty$, indicating that the RG flows 
to the strong-coupling fixed point with complete Kondo screening.
Thus, the Kondo scale, defined in this way, is generally finite even though 
$g\to \infty$. By equivalence of both Kondo sites, it is the same as the 
Kondo temperature $T_K$ appearing in the local spin susceptibility $\chi_f$ on 
the second Kondo site (see above).   
This yields the defining equation for $T_K$ \cite{Nejati16},  
\bEq
\ln \left(\frac{T_K(y)}{T_K(0)}\right) &= 
-y \, \alpha\, g_0^2 \, \frac{D_0}{T_K(y)} ~,
\label{eq:TKy}
\eEq
where $D_0/T_K(0)\gg 1$ was used and the coefficient 
$\alpha=\ln (\sqrt{2}+1)$ arises from fixing the integration constant 
by the condition $D(g\to\infty)=T_K(y)$.
Note that this is an implicit or selfconsistent equation, 
because the $\beta$-function itself [right-hand side of \Eq{eq:RG_eq}] 
depends parametrically on $T_K$ via $\chi_f$. 
The solution of \Eq{eq:TKy} also indicates that in the 2IK system the 
Kondo scale is modified by the RKKY interaction \cite{Nejati16}, 
i.e., $T_K=T_K(y)$.
In the absence of the RKKY interaction, the bare Kondo coupling, 
$g_0 \equiv g(D_0)$, satisfies
\bEq
- \frac{1}{2 g_0} = \ln\left( \frac{T_K(0)}{D_0} \right) ~.
\eEq
Hence, by defining a rescaled Kondo temperature, 
$ \tau(y) := \frac{T_K(y)}{T_K(0)}$, 
one obtains the equation for the RKKY-modified Kondo scale as
\bEq
\label{eq:tau}
\tau(y) \equiv \frac{T_K(y)}{T_K(0)}
= {\rm e}^{ -y\,\alpha\, g_0^2 \frac{D_0}{T_K(0)} \, \tau(y) } ~. 
\eEq
It can be solved in terms of the Lambert $W$ function,
\footnote{Indeed, the first branch of the Lambert function, $W_0$, is the 
solution. More information can be found in~\Ref{Veberic12}.
}
\bEq
\label{eq:TKsol}
\tau (y) = \frac{- \gamma_0 \, y} {W( -\gamma_0 \, y )} ~,
\eEq
with the parameter,

\bEq
\label{eq:gamma0}
\gamma_0 = g_{0}^2\,\alpha\,\frac{D_0}{T_K(0)} = 
\frac{\alpha}{4 (\ln(T_{K}(0)/D_0))^2} \frac{D_0}{T_{K}(0)} ~.
\eEq

\noindent
The solution $\tau(y)$ is shown in~\Fig{fig:TKy}, left panel \cite{Nejati16}.
It is seen that the solution of \Eq{eq:TKsol} ceases to exist when 
$y$ exceeds the critical strength, $y>y_c=1/({\rm e}\gamma_0)$, indicating
that the RG flow does not diverge, i.e. a strong coupling Kondo singlet
is not formed for $y>y_c$. As seen from \Eq{eq:gamma0}, the critical strength 
$y_c$ is determined alone by the single-impurity Kondo scale 
$T_K(0)$. At the breakdown point, the RKKY-induced suppression ratio 
of the Kondo scale takes the universal, finite value 
$T_K(y_c)/T_K(0)=1/{\rm e}\approx 0.368$, where {\rm e} is Euler's 
constant \cite{Nejati16}. We emphasize that, since the solution 
$1/{\rm e}\leq T_K(y)/T_K(0)\leq 1$ for all values of $1\geq y\geq 0$,
the term in brackets on the right-hand side of the RG equation (\ref{eq:RG_eq}) 
($\beta-$function) remains always positive and less than one, i.e.,
the RG flow remains always in the perturbatively controlled regime.

\section{Comparison with experimental results and discussion}\label{sec:results}

Having adapted the RG formalism of \Ref{Nejati16} to the Kondo suppression 
in a 2IK system RKKY-coupled by a 1D Cu chain, we are now in a position to 
compare with the corresponding experiments of \Ref{Neel11}. All spatial 
dependence is cast in the distance dependence of the RKKY parameter $y(n)$,
\Eq{eq:RKKYparameter}. Since the magnetic moment of a Co impurity atom is 
spread over a Co $d-$orbital occupying about one unit cell, we average 
$y(n)$ over one lattice spacing $a$ of the underlying Cu lattice 
(green dashed curve in \Fig{fig:TKy}, right panel) and insert the 
result into \Eq{eq:TKsol} 
to calculate the RKKY-induced Kondo suppression ratio $T_K(y(n))/T_K(0)$
as a function of the Cu chain length.
Since the RKKY wave number is fixed to $2k_F=\pi/a$ (half band filling, see 
section \ref{sec:RKKY}), this result has a single scale factor,
$\gamma_0$, as the only adjustable parameter to fit the experimental results. 
We adjust it to fit the experimental data for $n=2$. 
Note that, in fact, the {\it relative} suppression, $T_K(y(n))/T_K(0)$, 
with respect to the single-ion value of experimentally $T_K(0)=110$~K 
\cite{Neel11} has no adjustable parameter at all. As seen in 
\Fig{fig:TKy} (right panel), 
the agreement with the experimental data for $n\geq 2$ is quantitatively 
very good, with only a slight deviation for a Cu chain length of $n=6$.
It is remarkable that the 1D infinite chain model reproduces, 
without fit parameter, the periodicity as well as the relative amplitude 
of the $T_K(y(n))$ suppression accurately even for short chain lengths 
down to $n=2$, considering the crudeness of this model. 
It is important to note that our theory 
explains that the RKKY interaction of either sign 
(ferro- or antiferromagnetic) suppresses the Kondo scale, 
as is physically expected and experimentally 
observed \cite{Neel11}. Only for the shortest Cu chain, $n=1$, the
theory does not fit the experiment. We attribute this to the fact that 
the simple infinite-chain model fails for this very short cluster, because 
the local DOS and, hence, the RKKY interaction are strongly influenced
by boundary effects, the geometry of the wave functions and chemical bonds. 
More precise DFT calculations should be employed to calculate the 
RKKY coupling and the $y$-parameter in this case.  
Interestingly, however, the strongest experimentally observed suppression ratio
is $ T_K(y(n = 1))/T_K(0) = 46 / 110 \approx 0.42 $ 
(\Fig{fig:TKy}, right panel), which is remarkably close to the 
theoretical prediction $1/{\rm e}$, considering that the impurity 
spacing cannot be changed continuously in the experiment.

In summary, we have given a short review of the recently developed 
renormalization group method to tackle the influence of the RKKY 
interaction on the Kondo screening in multi-impurity and lattice 
Kondo systems \cite{Nejati16}. The application of this theory to 
the oscillation and suppression of the Kondo scale in Co-Cu$_n$-Co
chains \cite{Neel11} yield generally good, quantitative agreement. 
This further supports 
the validity of this RG method. The deviation of the theory 
from the experiment for the shortest chain, $n=1$, 
is attributed to boundary effects not included in the infinite chain model and
may be improved by DFT calculations of the input parameters of the RG.

\section*{Acknowledgements}
The authors are thankful for discussions with J. Kr\"{o}ger and N. N\'eel 
regarding the experimental results. This work was supported in part by 
DFG through SFB-TR 185.


\section*{References}
\bibliographystyle{iopart-num}
\bibliography{2IK_chain}

\providecommand{\newblock}{}
\begin{thebibliography}{10}
\expandafter\ifx\csname url\endcsname\relax
  \def\url#1{{\tt #1}}\fi
\expandafter\ifx\csname urlprefix\endcsname\relax\def\urlprefix{URL }\fi
\providecommand{\eprint}[2][]{\url{#2}}

\bibitem{Nejati16}
Nejati A, Ballmann K and Kroha J 2016 {\em arXiv:1612.03338\/}

\bibitem{Neel11}
N\'{e}el N, Berndt R, Kr\"{o}ger J, Wehling T~O, Lichtenstein A~I and
  Katsnelson M~I 2011 {\em Phys. Rev. Lett.\/} {\bf 107} 106804

\bibitem{Hewson97}
Hewson A~C 1997 {\em The Kondo Problem to Heavy Fermions\/} (Cambridge:
  Cambridge University Press)

\bibitem{Coleman07}
Coleman P 2007 {\em Handbook of Magnetism and Advanced Magnetic Materials\/}
  (Hoboken, NJ: Wiley) chap Heavy fermions: Electrons at the edge of magnetism

\bibitem{deHaas34}
De~Haas W~J, de~Boer J and van~den Berg G~J 1934 {\em Physica\/} {\bf 1} 1115

\bibitem{Ehm07}
Ehm D, H\"ufner S, Reinert F, Kroha J, W\"olfle P, Stockert O, Geibel C and von
  L\"ohneysen H 2007 {\em PRB\/} {\bf 76} 045117

\bibitem{GoldhaberGordon98}
Goldhaber-Gordon D, Shtrikman H, Mahalu D, Abusch-Magder D, Meirav U and
  Kastner M~A 1998 {\em Nature\/} {\bf 391} 156

\bibitem{Weis98}
Schmid J, Weis J, Eberl K and Klitzing K 1998 {\em Physica B\/} {\bf 256-258}
  182

\bibitem{Kouwenhoven00}
van~der Wiel W~G W~G, de~Franceschi S, Fujisawa T, Elzerman J~M, Tarucha S and
  Kouwenhoven L~P 2000 {\em Science\/} {\bf 289} 2105

\bibitem{Grobis07}
Grobis M, Rau I~G, Potok R~M and Goldhaber-Gordon D 2007 {\em Handbook of
  Magnetism and Advanced Magnetic Materials\/} (Hoboken, NJ: Wiley) chap The
  Kondo effect in mesoscopic quantum dots

\bibitem{Nygaard00}
Nygard J, Cobden D~H and Lindelof P~E 2000 {\em Nature\/} {\bf 408} 342

\bibitem{Park02}
Park J, Pasupathy A~N, Goldsmith J~I, Chang C, Yaish Y, Petta J~R, Rinkoski M,
  Sethna J~P, D H, McEuen P~L and Ralph D~C 2002 {\em Nature\/} {\bf 417} 722

\bibitem{Scott10}
Scott G~D and Natelson D 2010 {\em ACS Nano\/} {\bf 4} 3560

\bibitem{Ruderman54}
Ruderman M~A and Kittel C 1954 {\em Phys. Rev.\/} {\bf 96} 99

\bibitem{Kasuya56}
Kasuya T 1956 {\em Prog. Theor. Phys.\/} {\bf 16} 45

\bibitem{Yosida57}
Yosida K 1957 {\em Phys. Rev.\/} {\bf 106} 893

\bibitem{vVleck62}
van Vleck J~H 1962 {\em Rev. Mod. Phys.\/} {\bf 34} 681

\bibitem{Jones88}
Jones B~A, Varma C~M and Wilkins J~W 1988 {\em Phys. Rev. Lett.\/} {\bf 61} 125

\bibitem{Jones89}
Jones B~A and Varma C~M 1989 {\em Phys. Rev. B\/} {\bf 40} 324

\bibitem{Affleck95}
Affleck I, Ludwig A~W~W and A J~B 1995 {\em Phys. Rev. B\/} {\bf 52} 9528

\bibitem{Mitchell15}
Mitchell A~K, Derry P~G and Logan D~E 2015 {\em Phys. Rev. B\/} {\bf 91} 235127

\bibitem{Craig04}
Craig N~J, Taylor J~M, Lester E~A, Marcus C~M, Hanson M~P and Gossard A~C 2004
  {\em Science\/} {\bf 304} 565

\bibitem{Sasaki06}
Sasaki S, Kang S, Kitagawa K, Yamaguchi M, Miyashita S, Maruyama T, Tamura H,
  Akazaki T, Hirayama Y and Takayanagi H 2006 {\em Phys. Rev. B\/} {\bf 73}
  161303(R)

\bibitem{Jeong01}
Jeong H, Chang A~M and Melloch M~R 2001 {\em Science\/} {\bf 293} 2221

\bibitem{Chang09}
Chang A~M and Chen J~C 2009 {\em Rep. Prog. Phys.\/} {\bf 72} 096501

\bibitem{Chen99}
Chen W, Jamneala T, Madhavan V and Crommie M~F 1999 {\em Phys. Rev. B\/} {\bf
  60} R8529

\bibitem{Bork11}
Bork J, Zhang Y~H, Diekh{\"o}ner L, Borda L, Simon P, Kroha J, Wahl P and Kern
  K 2011 {\em Nature Phys.\/} {\bf 7} 901

\bibitem{Prueser14}
Pr\"user H, Dargel P~E, Bouhassoune M, Ulbrich R~G, Pruschke T, Lounis S and
  Wenderot M 2014 {\em Nature Comm.\/} {\bf 5} 5417

\bibitem{Uhsaghy00}
Ujs\'aghy O, Kroha J, Szunyogh L and Zawadowski A 2000 {\em Phys. Rev. Lett.\/}
  {\bf 85} 2557

\bibitem{Aristov97}
Aristov D~N 1997 {\em Phys. Rev. B\/} {\bf 55} 8064

\bibitem{Economou06}
Economou E~N 2006 {\em Green's functions in quantum physics\/} (Berlin:
  Springer)

\bibitem{Andrei83}
Andrei N, Furuya K and Lowenstein J~H 1983 {\em Rev. Mod. Phys.\/} {\bf 55} 331

\bibitem{Veberic12}
Veberi\'c D 2012 {\em Computer Phys. Comm.\/} {\bf 183} 2622; arXiv:1209.0735

\end{thebibliography}


\end{document}